\documentclass[preprintnumbers,superscriptaddress,pra]{revtex4}
\usepackage{amssymb}
\usepackage{amsmath}
\usepackage{epsfig}
\usepackage{graphicx}
\usepackage{color}

\input{tcilatex}
\begin{document}

\title{Comment on \textquotedblleft New General Relativistic Contribution to
Mercury's Perihelion Advance\textquotedblright }
\author{J. Y. Zhou}
\affiliation{School for Theoretical Physics, School of Physics and Electronics, Hunan
University, Changsha 410082, China}
\author{Z. Li}
\affiliation{School for Theoretical Physics, School of Physics and Electronics, Hunan
University, Changsha 410082, China}
\author{Q. Li}
\affiliation{School for Theoretical Physics, School of Physics and Electronics, Hunan
University, Changsha 410082, China}
\author{X. Wang}
\affiliation{School for Theoretical Physics, School of Physics and Electronics, Hunan
University, Changsha 410082, China}
\author{Q. H. Liu}
\email{quanhuiliu@gmail.com}
\affiliation{School for Theoretical Physics, School of Physics and Electronics, Hunan
University, Changsha 410082, China}
\affiliation{Synergetic Innovation Center for Quantum Effects and Applications (SICQEA),
Hunan Normal University,Changsha 410081, China}
\date{\today }

\begin{abstract}
C. M. Will in a recent Letter predicts a new Mercury's precession rate which
is $100$ times larger than the second-post-Newtonian contribution, our
calculation is about $30$ times larger. Thus the new Mercury's precession
rate is about $6.4\ast 10^{-8}$, instead of $6.2\ast 10^{-8}$, degrees per
century.
\end{abstract}

\maketitle

In a recent Letter, C. M. Will \cite{1} reported a new general relativistic
contribution to the perihelion advance of Mercury, arising in part from
relativistic \textquotedblleft crossterms\textquotedblright\ in the
post-Newtonian equations of motion between Mercury's interaction with the
Sun and with the other planets, and in part from an interaction between
Mercury's motion and the gravitomagnetic field of the moving planets. These
previously ignored effects is likely to be detectable by the BepiColombo
mission to place and track two orbiters around Mercury, scheduled for launch
around 2018. This Comment does not alter the main result/conclusion but
presents a correct comparison between the new predicted result and the
second order correction to the perihelion advance of Mercury within
Schwarzschild metric in Harmonic coordinates that Will \cite{1} uses.

The Schwarzschild metric in Harmonic coordinates predicts the perihelion
advance of Mercury, up to the second order of perturbation parameter $b^{2}$
is, from the singular perturbation theory \cite{2},%
\begin{equation}
\Delta \varphi \approx 2\pi (3b^{2}+\frac{3}{4}(34+e^{2})b^{4}),  \label{1}
\end{equation}%
where $e$ denotes the eccentricity of Mercury's orbit, and $b^{2}$ is
defined by,%
\begin{equation}
b\equiv \frac{GM}{c^{2}a(1-e^{2})},  \label{2}
\end{equation}%
where $G$ is Newtonian gravitational constant, $M$ is the mass of the Sun, $c
$ is the velocity of the light, and $a$ is the semi-major axis of the
Mercury's orbit. The first term $6\pi b^{2}$ is the celebrated perihelion
advance of Mercury \cite{1}, and the second term is the second order
correction,%
\begin{equation}
\Delta _{2}\varphi =\frac{3\pi }{2}b^{4}(34+e^{2}).  \label{3}
\end{equation}%
However, the 2PN term of in Table 1 of \cite{1} gives, where notation $%
GM/(c^{2}p)$ is in fact $b^{2}$, 
\begin{equation}
\Delta _{2}\varphi ^{\prime }=-\frac{3\pi }{2}b^{4}(10-e^{2}).  \label{4}
\end{equation}%
The ratio of magnitude of $\left\vert \Delta _{2}\varphi ^{\prime }/\Delta
_{2}\varphi \right\vert $ gives, with $e=0.20563$ \cite{3},%
\begin{equation}
\frac{(34+e^{2})}{(10-e^{2})}\approx 3.4168  \label{5}
\end{equation}%
This result depends on an approximate value of the eccentricity only,
without reference to other parameters.

The second-post-Newtonian expression in \cite{1} is different from ours 
\textit{both in sign and in magnitude}, and the new Mercury's precession
rate predicted by \cite{1} is not $100$ times larger than the
second-post-Newtonian contribution in magnitude, but about only $30$ times
larger. To note that predicted new Mercury's precession rate is about $%
6.2\ast 10^{-8}$ degrees per century. With considering about $34\%$ increase
from the second-post-Newtonian contribution, calculations show that the
result should be $6.4\ast 10^{-8}$ degrees per century.

\begin{acknowledgments}
This work is financially supported by National Natural Science Foundation of
China under Grant No. 11675051.
\end{acknowledgments}

\end{document}